\documentclass[fleqn,usenatbib]{mnras}

\usepackage{newtxtext,newtxmath}
\usepackage[T1]{fontenc}

\DeclareRobustCommand{\VAN}[3]{#2}
\let\VANthebibliography\thebibliography
\def\thebibliography{\DeclareRobustCommand{\VAN}[3]{##3}\VANthebibliography}

\usepackage{color}

\usepackage{graphicx}	% Including figure files
\usepackage{amsmath}	% Advanced maths commands
\newcommand\secref[1]{Sect.~\ref{#1}}

\newcommand\figref[1]{Fig.~\ref{#1}}

\newcommand\equref[1]{Eq.~\eqref{#1}}

\newcommand\tabref[1]{Table~\ref{#1}}

\title[Raining Rocks]{Raining Rocks: An analytical formulation for collision timescales in planetary systems}

\author[S. Torres et al.]{
Santiago Torres,$^{1,2}$\thanks{E-mail: storres@astro.ucla.edu}
Smadar Naoz,$^{1,2}$
Gongjie Li$^{3}$
and Sanaea C. Rose$^{1,2}$
\\
$^{1}$Department of Physics and Astronomy, University of California, Los Angeles, CA 90095, USA\\
$^{2}$Mani L. Bhaumik Institute for Theoretical Physics, University of California, Los Angeles, CA 90095, USA\\
$^{3}$Center for Relativistic Astrophysics, School of Physics, Georgia Institute of Technology, Atlanta, GA 30332, USA
}

\date{Accepted XXX. Received YYY; in original form ZZZ}

\pubyear{2022}

\begin{document}
\label{firstpage}
\pagerange{\pageref{firstpage}--\pageref{lastpage}}
\maketitle

%---------------------------------------------%
%------------------------------------------- Abstract ---------------------------------------------%

\begin{abstract}
The dynamical interaction of minor bodies (such as comets or asteroids) with planets plays an essential role in the planetary system's architecture and evolution. As a result of these interactions, structures like the Kuiper belt and the Oort cloud can be created. In particular, the collision of minor bodies with planets can drastically change the planet's internal and orbital evolution. We present an analytic formulation to determine the collision timescale for a minor body to impact a planet for arbitrary geometry. By comparing with a suite of detailed N-body simulations and an analytical method for collision timescales in the solar system, we confirmed the accuracy of our analytic formulation. As a proof of concept, we focused on the collision rate of minor bodies randomly distributed around a Jupiter-like planet, emulating a Kuiper belt-like disk. We show that our analytical method yields in good agreement with the numerical simulations. The formalism presented here thus provides a succinct and accurate alternative to numerical calculations.
\end{abstract}


%--------------------------------------------------%
%--------------------------------------------- Intro ---------------------------------------------------%
\section{Introduction}
\label{intro}

Minor bodies such as comets and asteroids in the solar system are remnants of the planet formation process \citep{Kokubo2002,Kenyon2006,Wyatt2008,Johansen2017}.  These objects play an important role in the evolution of their planetary system \citep{Nesvorny2018,Torres2019,Cai2019a}. In particular, the dynamical evolution of these bodies in any planetary system is dominated by the gravitational interaction with major bodies such as the planets. As the comets come close to planetary regions, planets become the main influence for these objects.

Thus, gravitational interactions with planets, such as close encounters and collisions, may have influenced the planets' history, composition, structure, and evolution \citep{Asphaug2006,Brasser2020,Morgan2021}. Examples of these processes include cometary impacts that may be responsible for the dawn-dusk asymmetry of Mercury’s exosphere \citep[e.g.,][]{Benz1988,Pokorny2017}, the changes of the surface and atmosphere on Mars \citep[e.g.,][]{Carr1989,Melosh1989,Woo2019}, and the dynamical evolution of the gas giants and trans-neptunian objects (\citep{Gomes2005,MunosGutierrez2021}, for more examples see \cite{Stern1995,MarovRickman2001,Charnoz2003}. Furthermore, collisions with these remnants may have a dramatic effect on a planet's orbit. For example, repeated collisions may have resulted in the tilt of Uranus \citep[e..g,][]{Brunini1995,Parisi1997,Rogoszinski2021}. Lastly, the Chicxulub impact on Earth is suspected to be the main cause of the extinction of the dinosaurs \citep[e.g.,][]{Alvarez1980,Schulte2010}.

Cometary (and other minor body) impacts in the solar system have been extensively studied in the literature \citep[e.g.,][]{Opik1951,Kessler1981,Greenberg1988,Bottke1993,MarovRickman2001,Muinonen2001,Valsecchi2005,Rickman2014a}. Of particular interest is the impact rate of comet collisions with planets. These calculations are often based on the \"Opik's analytic method \citep{Opik1951}.  However, these methods are often tuned to model cometary impacts in the inner part of the solar system. As a result, it is challenging to calculate collision rates in other planetary systems with different architectures, arbitrary configurations and geometries than the solar system. 

Here we present a succinct and accurate model to calculate the collisions rate and timescale for a minor body to impact a planet. Our methodology is applicable for all geometries and configurations and is consistent with direct numerical calculations. In \secref{sec2}, we present our model for collisional timescales. In \secref{sec3} we test our model by comparing our predictions with a well known analytic method for collision rates in the solar system and detailed N-body simulations. Finally, we discuss our results in \secref{discussion}.

%---------------------------------------------------%
%---------------------------------------------- Sec 2
%---------------------------------------------------%
\section{Collisional timescales for particle impacts on planets}
\label{sec2}

Here we present a general analytical approach, to calculate the collision rate of a minor body with a planet  for arbitrary geometry of interaction. Hereafter we refer to a minor body as a particle to highlight the wide range of application.

\begin{figure}
\centering
\includegraphics[width=1.\columnwidth]{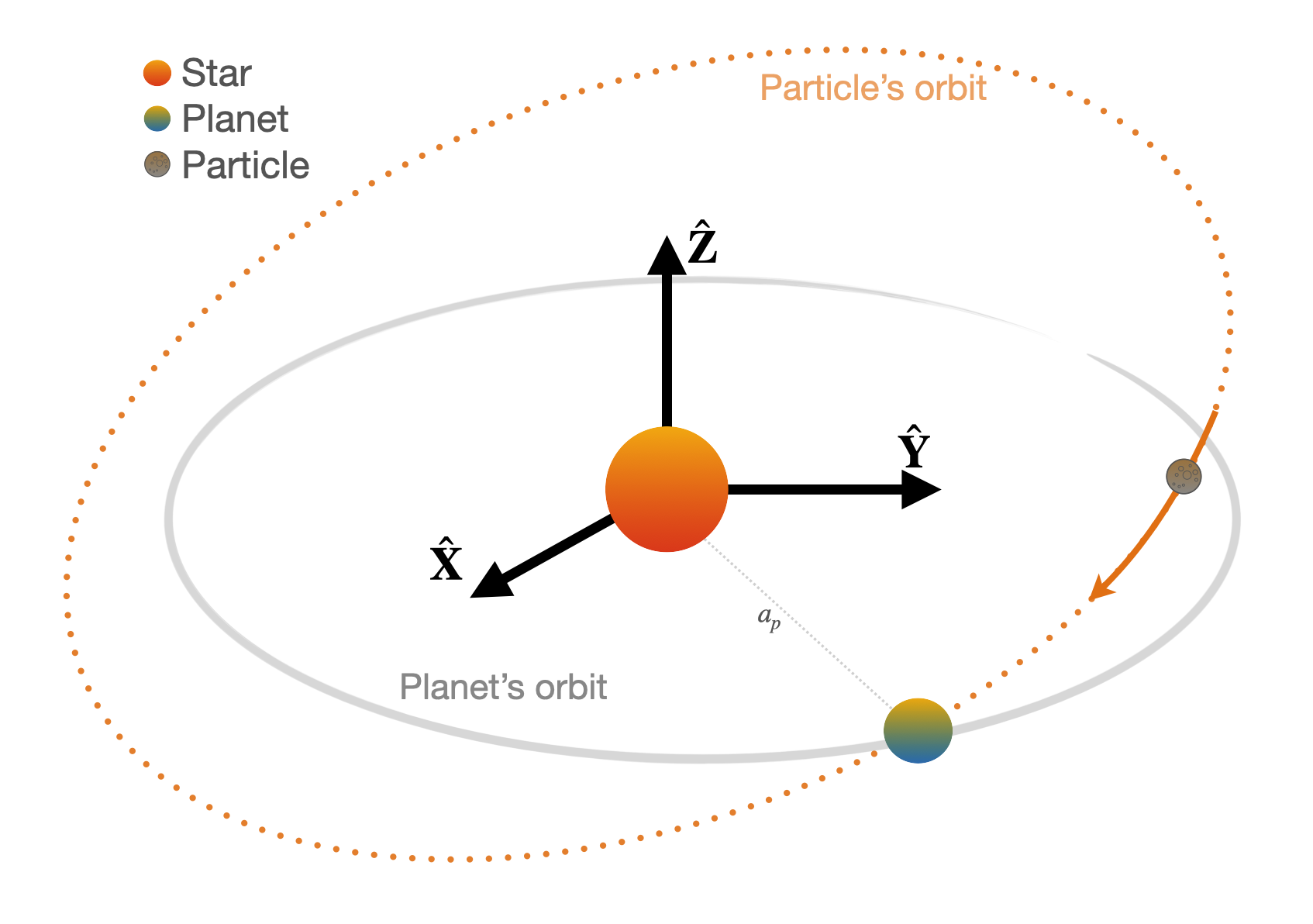}
\caption{Orbital motion of a particle  about it's host star with respect to the planet's orbital plane. The semi-major axis of the planet is given by $a_p$.} 
\label{geometry}
\end{figure}

Consider the collision rate, $\Gamma_{\rm coll}$ (\equref{eq:Gamma}), of a particle with a planet.  This rate can be calculated by assuming a population of planet-orbital-crossing particles a with number density $n$ that will eventually collide with the planet. The relative velocity between the planet and the particle ($v_{\rm rel}$), and the cross section of interaction ($\sigma$). We assume an arbitrary configuration for the planet and particle; see \figref{geometry} for illustration. Thus, the collision rate can be approximated as \citep[e.g.,][]{BinneyTremaine2008,Nesvorny2020,Rose2020}
%%-------------------- Gamma
\begin{equation}\label{eq:Gamma}
\Gamma_{\rm coll} = n\,v_{\rm rel}\,\sigma.
\end{equation}

%%------------------- Geometry of the encounter Based on Fig. 2.13 

%%%-------------------- Number density and volume. 
The number density is then simply $n=N_c/V$, where $N_c$ is the number of colliding particles and $V$ is the volume where collision can take place, and is given by,
\begin{equation}\label{volume}
V= 2\pi^{2}a_{p}R_{p}^{2}\,sin\,i_{c},
\end{equation}
where  $a_{p}$ and $R_{p}$ are the semi-major axis and radius of the planet respectively and $i_{c}$ is the inclination of the particle.

%%---------------------- Relative velocity
The relative velocity magnitude between the particle and the planet is given by  the magnitude difference between the velocity vector of the planet ${\bf v}_p$ and the particle ${\bf v}_c$, in other words, $v_{\rm rel}=| {\bf v}_{\rm rel}| =| {\bf v}_p -{\bf v}_c |$. To obtain the velocity vectors of the planet and the particle, we first need to establish the geometry of the encounter. We considered a reference plane in which the planet is at the center. The orbital motion of the particle about the star with respect to the center is in three-dimensional space.

The position vector of the particle in the frame of it's bound orbit about the host star is: ${\bf r}_c $=$x\,\hat{\mathbf{x}}+ y\,\hat{\mathbf{y}}+ z\,\hat{\mathbf{z}}$, where $x=r_{c}\,cos\,f_{c}, y=r_{c}\,sin\,f_{c}, z=0$. Where $f_c$ is the true anomaly of the particle, and $r_c$ is given by: 
\begin{equation}\label{rc}
r_{c}= a_{c}\,\frac{1-e_{c}^{2}}{1+e_{c}\,\cos\,f_{c}} \ . 
\end{equation}

Then, the position vector projected to the plane of the planet is given by ${\bf R}_{1}{\bf R}_{2}{\bf R}_{3} {\bf r}_c$, where ${\bf R}_j$, $j=1,2,3$ are the rotation matrices given by \citep[e.g.,][]{MurrayDermott2000},
%\begin{equation}
\begin{center}
\begin{eqnarray}
R_{1}=
\begin{pmatrix}
cos\,\omega  & -sin\,\omega & 0 \\
sin\,\omega & cos\,\omega & 0 \\
0 & 0 & 1
\end{pmatrix}\ \ ,
R_{2} =
\begin{pmatrix}
1 & 0 & 0 \\
0 & cos\,i  & -sin\,i\\
0 & sin\,i & cos\,i
\end{pmatrix}\ \ ,\nonumber\\
R_{3}=
\begin{pmatrix}
cos\,\Omega  & -sin\,\Omega & 0 \\
sin\,\Omega & cos\,\Omega & 0 \\
0 & 0 & 1
\end{pmatrix}. \ 
\end{eqnarray}
\end{center}
Therefore, the components of the position vector of the particle projected on the planet's can be calculated by,
\begin{equation}
 \begin{pmatrix}
 X\\
 Y\\
 Z
  \end{pmatrix}
=
R_{1}R_{2}R_{3}
 \begin{pmatrix}
 x\\
 y\\
 z
  \end{pmatrix}
\label{vector}\end{equation}

Consequently, the velocity vectors of the particle ($\mathbf{v}_{c}$) and the planet ($\mathbf{v}_{p}$) in their own individual orbital planes at any give time are given by
\begin{equation}\label{vc}
\mathbf{v}_{c,p}= \left( - \frac{h_{c,p}\,\sin\,f_{c,p}}{a_{c,p}\,(1-e_{c,p}^{2})},\frac{h_{c,p}\,(e_{c,p}+\cos\,f_{c,p})}{a_{c,p}(1-e_{c,p}^2)}, 0\right) \ ,
\end{equation}
where, the subscript $c$ and $p$ stands for the particle and the planet respectively. The specific angular momentum of the particle (planet) $h_{c}$ ($h_p$) is given by,
\begin{equation}\label{hc}
h_{c,p}=\sqrt{G\,(M_s + M_p)\,a_{c,p}\,(1-e_{c,p}^2)} \ ,
\end{equation}
where  $a_c~(a_{p})$, $e_c~(e_{p})$ and $f_c~(f_{p})$, are the semi-major axis, eccentricity and true anomaly of the particle (planet).  Additionally, $M_s$ and $M_p$ are the mass of the star and the planet respectively. Thus, the relative velocity is given by:
\begin{equation}
v_{\rm rel}=|{\bf v}_p -R_{1}R_{2}R_{3} {\bf v}_c| \ .
\label{velrel}\end{equation}
Where recall that we are rotating the particle velocity vector to the planet frame using $R_{1}R_{2}R_{3}$. The relative velocity between the particle and the planet is calculated at the moment of collision. The collision of the particle with the planet takes place at the node, where the cometary orbital plane and the planet's plane coincide.

%%%------------------- cross section
Finally, the cross section $\sigma$, enhanced by gravitational focusing, is given by:
\begin{equation}\label{eq:sigma}
\sigma=\pi\left (R_p^2+R_p\,\frac{2\,G\,M_p}{v_{\rm rel}^2}\right) \ ,
\end{equation}
Substituting equations \ref{volume}, \ref{velrel}, and \ref{eq:sigma} into \equref{eq:Gamma} we obtain a final expression for the collision rate per year as a function of the particle's orbital elements $\Gamma_{\rm coll}(a_{c},e_{c},i_{c},\Omega_c,\omega_c,f_c)$:

\begin{equation}\label{gamma}
\Gamma_{\rm coll} =\frac{N_c}{2\pi^{2}a_{p}R_{p}^{2}\,sin\,i_{c}}\,\left (v_{\rm rel}\,\pi\,R_p^2+\pi\,R_p\,\frac{2\,G\,M_p}{v_{\rm rel}}\right)
\end{equation}

Equation \ref{gamma} represents the most general expression for a collision between a minor body and a planet in an arbitrary geometry.

%---------------------------------------------------
%--------------------------------------------- Sec 3
%---------------------------------------------------
\section{Comparison with analytic and numerical methods}
\label{sec3}

In this section we tested our model by comparing it with the often use \"Opik method for collisions in the solar system (see also Appendix \ref{opikvscoll}) and with detailed N-body simulations (Sec.\ref{comparison}).

%%%%%%%%%%%%%%%%%%%%%%%% Opik %%%%%%%%%%%%%%%%%%%%% 
%%%%%%%%%%%%%%%%%%%%%%%%%%%%%%%%%%%%%%%%%%%%%%%%%%%
\subsection{The \"Opik Method} 
\label{sec3.1}

The \"Opik method \citep{Opik1951} in its original form provides an expression for the collision rate of particles (asteroids or comets) with planets. In this 1951 method, the planet is assumed to be fixed in space in a circular orbit and the colliding particle on an arbitrary orbit. The collision happened when the orbit of the two bodies intersect. The \"Opik method assumes a restricted 3-body problem, considering the small body massless and moving on an unperturbed heliocentric Keplerian orbit.

The \"Opik method considers two main parts to calculate the collision rate:  the relative velocity between the particle and the planet and the collisional area or cross-section. The \"Opik method often uses units $G=1$ and assumes the star's mass $M_{s}$=1\,$M_{\odot}$.

The reference frame is set so that the particle is at one of the nodes of its orbit when the encounter with the planet occurs. Therefore, the relative velocity $U_{\rm opik}$ can be expressed in terms of the Tisserand parameter with respect to the planet and is given by \citep[e.g.,][]{Opik1951,Carusi1990},
\begin{equation}
U_{\rm opik}=\sqrt{3-T} \ ,
\label{vel_T}\end{equation}
where T, is the Tisserand parameter which is defined as,
\begin{equation}
T =\frac{a_{p}}{a_{c}} + 2\,cos\,i_{c} \sqrt{ \frac{a_{c}}{a_{p}} (1-e_{c}^{2})} \ .
\label{T}\end{equation}
The relative velocity components, $U_x$, $U_y$, and $U_z$ are given by \citep[see e.g.,][]{Carusi1990}, 
\begin{eqnarray}
U_{x} &=& \pm\sqrt{2-1/a_{c} - a_{c}(1-e_{c}^{2})}, \nonumber\\
U_{y} &=& \sqrt{a_{c}(1-e_{c}^{2})}\,cos\,i_{c} -1,\nonumber\\
U_{z} &=& \pm\sqrt{a_{c}(1-e_{c}^{2})}\,sin\,i_{c}\,.
\label{velxyz}\end{eqnarray}

%%% cross section

When a particle reaches the Hill's sphere of a planet, the Sun's perturbations can be neglected, and the trajectory of the particle can be modeled as a planetocentric. Once inside of the Hill's region, a particle can collide with the planet if the pericenter distance of the particle $q_{c}$ is smaller or similar to the radius of the planet $R_{p}$, i.e., $q_{c}\leqslant R_{p}$. Therefore the the cross-section for interaction can be expressed by $\sigma_{\rm opik}$ \citep[e.g.,][]{Opik1951,Opik1976},
\begin{equation}
 \sigma_{\rm opik}= R_{p}\,\sqrt{1+\frac{2\,G\,M_{p}}{{U_{\rm opik}}^{2}\,R_{p}}}\ .
 \label{eq:sigmaopik}
\end{equation}

With the expression for the encounter velocity $U_{\rm opik}$ and the cross-section $\sigma_{\rm opik}$, the collision rate per year as a function of the particle's orbital elements $\Gamma_{\rm opik}(a_{c},e_{c},i_{c})$ is calculated as followed.

 The \"Opik method considers a particle in a heliocentric orbit, which crosses two times an sphere of radius $r=1$. The radial velocity of the particle is $U_{x}$. Then the time spent of the particle in the sphere is $dt=2\,dr/U_{x}$. The number of particles $N_c$ in the sphere per orbital revolution can be calculated as $N_c=dt/P$, where $P$ is the particle's orbital period. \cite{Opik1951} showed that a particle could only be found inside a band with two parallel latitudes $\pm i$ and a volume $dV=4\pi\,sin\,i\,dr$. Thus, the collision rate of a particle can be expressed by \citep[see e.g.,][for a detailed derivation]{Opik1976},
\begin{equation}
\Gamma_{\rm opik} =\frac{\sigma_{\rm opik}^{2}\,U_{\rm opik}}{\pi\,sin\,i_c |U_{x}|}.
 \label{pc}
\end{equation}

Despite the simplicity, the \"Opik's method yields consistent results for Jupiter-family comets \citep[e.g.,][]{Greenberg1988,Nakamura1998,Dones1999}. However, it fails to accurately model the collision rate for those particles with Tisserand parameter $T\geqslant3$. 

Many improvements in \"Opik original theory have been done by several authors \citep[e.g.,][]{Nakamura1998,Manley1998,Dones1999,Levison2000,Zahnle2001,Vokrouhlicky2012,Pokorny2013,Rickman2014a,JeongAhn2017,Vokrouhlicky2019,Abedin2021}, creating a variety of \"Opik-like models that address some of the existing issues of the classic method. These \"Opik like-methods represent a quick (but at times less accurate) alternative to more robust numerical simulations. However, these expansions are mainly tuned for objects in the inner parts of the solar systems and they lack the flexibility to model minor bodies in exo-planetary systems for a wide range of configurations.

%%%%%%%%%%%%%%%%%%%%%%%%%%%%%%%%%%%%%%%%%%%%%%%%
%%  Nbody
%%%%%%%%%%%%%%%%%%%%%%%%%%%%%%%%%%%%%%%%%%%%%%%%

\subsection{Numerical Method}
\label{nbody}

We used the N-body package \texttt{REBOUND} \citep{Rein2012} with the WH-Fast integrator \citep{Rein2012} to calculate the collisional history of minor bodies with a planet. We considered a system compose by a solar mass star and a Jupiter-like planet with semi-major axis $a_{p}=5.2$\,au, eccentricity $e_{p}=0.05$, mass $M_{p}=0.001$\,$M_{\odot}$. We used an inflated collisional radius $R_{p}$=$(M_{p}/3M_\star)^{1/3}\,a_p$ (to assure more collisions in shorter time). We added $5000$ test particles representing the minor bodies. To compare the numerical simulation with the analytical calculation we construct two representative runs (see Appendix \ref{sim}). In one, {\bf R-inc}, we vary only the initial inclination but keep all of the other orbital parameters constant, and in the other, {\bf R-ecc} we vary only the initial eccentricity of the particle. The full set of initial conditions are described in \tabref{table1}. We model the collision of the particles as inelastic encounters. For simplicity, every particle that collided with the planet was removed from the simulation. The simulation was run up to $10^4$~yrs. We note that the number of collisions does not converge on this timescale. As a function of time, the number of particles that undergo collisions increases, as expected. We performed a series of tests using a simulation time of $10^5$~yrs, and we did not find qualitatively change our results. In Appendix\,\ref{sim} we show the results of the simulations. 

\begin{table}[h!]
\centering
\caption{Input orbital elements of the particle and planet in the numerical simulations: $a_c$, $e_c$, $i_c$, $\Omega_c$, $\omega_c$, and $f_c$, $f_p$ represents the semi-major axis, eccentricity, inclination, longitude of the ascending node, argument of periapsis, and true anomaly of the particle and the particle, respectively.}
\begin{tabular}{l|ccccccc} 
name & $e_c$ & $a_c$ & $i$ & $\Omega_c$& $\omega_c$ &$f_c$ &$f_p$ \\
     &     & [au] & [deg] & [deg] & [deg] &[deg] &[deg] \\
     \hline
  {\bf R-inc}   & 0.5 & 5.5 & $0-180$ & 0 & 0 & 0 & 0 \\
   {\bf R-ecc}  & $0-1$ & 5.5 & $27.5$ & 0 & 0 & 0 & 0 \\
%\bottomrule 
\label{table1}
 \end{tabular}
\end{table}

%%%%%%%%%%%%%%%%%%%%%%%%%%%%%%%%% Gamma vs Opik vs Nbody 
\subsection{Comparison with Analytic and Numerical Methods} 
\label{comparison}

\begin{figure}	
\includegraphics[width=1.\columnwidth]{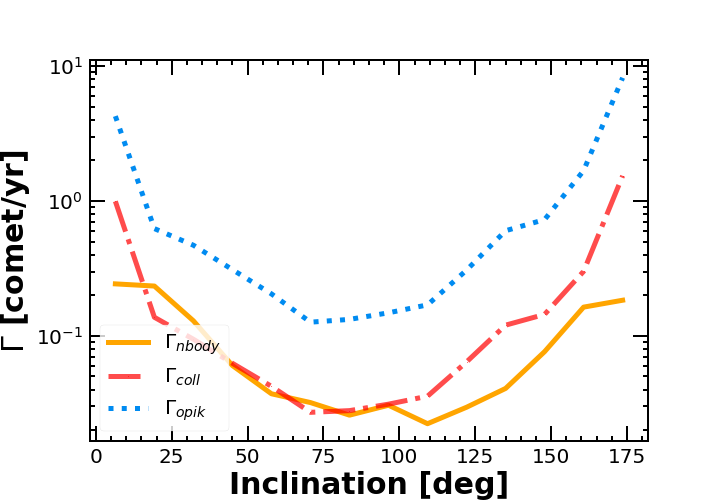}
\includegraphics[width=1.\columnwidth]{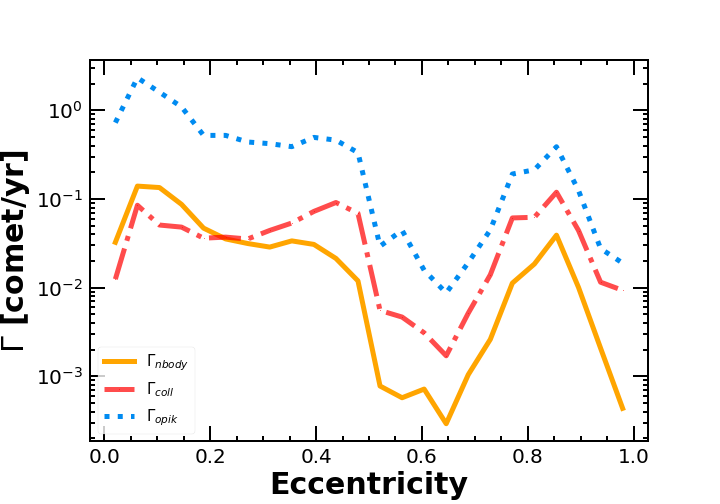}
\caption{Comparison between the N-body simulation (solid orange line), the \"Opik method (dotted blue line) and the $\Gamma_{\rm coll}$ (dash-dot red line). The top panel shows the collision rate as a function of the inclination using the values for {\bf R-inc} listed in Table\,\ref{table1}. While the bottom panel shows the distribution for {\bf R-ecc} and the collision rate as a function of the eccentricity.} 
\label{fig:nbody_gamma_opik}
\end{figure}

In \figref{fig:nbody_gamma_opik}, we show the average collision rate from the simulation, for different eccentricity and inclination bins (orange solid line). We compare the numerical result with the calculated $\Gamma_{\rm coll}$ and $\Gamma_{\rm opik}$ (red dash-dot and blue dotted lines, receptively). We note that in both cases we use the orbital parameters of the particles at the onset of collision. Before the  particles collided their orbit evolves as expected from three-body evolution \citep[e.g.,][]{Naoz2017}, thus, their initial conditions from Table \ref{table1} differs from their orbital parameters when they collide. 

As depicted in \figref{fig:nbody_gamma_opik}, our analytical rate calculation, $\Gamma_{\rm coll}$ is consistent with the N-body rate in both its functional form and value.

Note that $\Gamma_{\rm opik}$, is at times few orders of magnitude different than the numerical results.  Furthermore, as clearly seen in \figref{fig:nbody_gamma_opik}, $\Gamma_{\rm opik}$ estimated higher collision rate for circular orbits, at odds with the numerical and $\Gamma_{\rm coll}$ estimations.

%---------------------------------------------------
%--------------------------------------------------- Sec 4
%---------------------------------------------------

\section{Summary and Discussion} 
\label{discussion}
%%%%%%%%%%%%%%%%%%%%%%%%%%%%%%%%%
Here we present an analytical model to determine the collision rate of a minor body (particle) with a planet for any type encounter geometry and orbit (\equref{gamma}). We tested our formulation by comparing  with the \"Opik method (\secref{sec3.1}) and detailed N-body simulations (\secref{nbody}). As a proof of concept we choose two representative examples, one for which we vary the colliding particles eccentricities, and the other, by varying their inclinations. Our prediction for the collision rate of a particle impacting a planet is consistent with the simulations, but differ with the \"Opik method. 

The inconsistency between our model and the \"Opik method are mainly due to the singularities produced by the \"Opik method not present in our model. The \"Opik method fails to estimate the collision rate for those particles with small values of inclination and $|U_x|$, producing a singularity since \equref{opik_unit} goes to infinity (see \figref{tcollvspcoll}). Therefore bodies similar to the centaurs, nearly isotropic and long-period comets, can not be accurately model following \equref{opik_unit}. These objects are expected to be abundant in exo-planetary systems as a consequence of planet formation \citep[e.g.,][]{Wyatt2008,Johansen2017}. On the other hand, $\Gamma_{\rm coll}$ produces better estimations for the collision rate for all varieties of orbital elements. This is because we allow for arbitrary geometry. As a result, the function $\Gamma_{\rm coll}$ allows us to model any type of minor body orbits given the flexibility to estimate collisional rates in exo-planetary systems accurately. 

We note that the \"Opik-like methods might provide better estimations than the original one \citep[e.g.,][]{Valsecchi2005,JeongAhn2017,Vokrouhlicky2019,Abedin2021}. However, a detailed comparison with the variations of the method is beyond the scope of our paper. We omit these comparisons because our intention is not to adapt or extend the \"Opik theory for collisions to any exo-planetary system. Therefore, we focus on a simple comparison with the backbone of the theory, the classic \"Opik method \citep{Opik1976}.

Our formulation $\Gamma_{\rm coll}$ provides a succinct solution to determine the collision rate of a particle as a function of its orbital elements ($\Gamma_{\rm coll}(a_{c},e_{c},i_{c},\Omega_c,\omega_c,f_c)$). These allowed us to model the collision of particles with planets for any encounter geometry and orbit, providing an accurate alternative to costly N-body simulations.

%---------------------------------- acknowledgments
\section*{Acknowledgments}

ST expresses his gratitude to Ylva G\"otberg and Erez Michaely, for their helpful discussions and comments to the present work. ST, SN, GL thank NASA-ATP: AWD-000836-G1. Furthermore, ST and SN thank partial support from the NSF through grant No. AST-1739160 and Howard and Astrid Preston for their generous support. SR thank NASA-ATP grant number 80NSSC20K0505, as well as Nina Byers Fellowship and Michael A. Jura Memorial Graduate Award for support.

%%%%%%%%%%%%%%%%%%%%%%%%%%%%%%%%%%%%%%%%%%%%%%%%%%
\section*{Data Availability}

The python scripts used to generate the data for this work can be accessed here: https://santiago-torres.com/Research

%%%%%%%%%%%%%%%%%%%% REFERENCES %%%%%%%%%%%%%%%%%%
\bibliographystyle{mnras}
\bibliography{biblio} 

%%%%%%%%%%%%%%%%% APPENDICES %%%%%%%%%%%%%%%%%%%%%
\appendix

%%%%%%%%%%%%%%%%%%%%%%%%%%%%%%% %  Opik with units

\section{\"Opik units}
\label{opik_units}

Many studies, following \"Opik-like methods, adopted the Jacobi normalized units, in  which $G=1$, $M_s=1$, the heliocentric distance of the particle at the specified time $r=1$ and the mean motion $n=1$. In order to accurately  compare with our method and the N-body simulations, we have to bring back the proper units. The Tisserand parameter (Eq.\,\ref{T}) and the relative velocity (Eq.\,\ref{vel_T}) can be rewritten as: 
\begin{equation}
T = \frac{G M_s}{a_p} + 2\,GM_s\,\sqrt{\frac{a_c(1-e_{c}^{2})}{a_{p}^3}}\,cos\,i_{c}
\label{T_u}\end{equation}
and,
\begin{equation}
U_{\rm opik}^2 = n^2 a_p^2 + \frac{2 G M_s}{a_p} - T \ ,
\label{U_units}\end{equation}
where $n$ is the mean motion. The relative velocity components are given by,
\begin{eqnarray}
U_{x} &=& \pm\sqrt{G M_s\,\left (\frac{2}{r}-\frac{1}{a_{c}} - \frac{a_{c}(1-e_{c}^{2})}{r^2}\right)}\, , \nonumber\\
U_{y} &=& \sqrt{G M_s\,\left (\frac{a_{c}(1-e_{c}^{2})}{r}\right)}\,cos\,i_{c} -1\, ,\nonumber\\
U_{z} &=&\pm\sqrt{G M_s\,\left (\frac{a_{c}(1-e_{c}^{2})}{r}\right)}\,sin\,i_{c}\,.
\label{Uxyz}\end{eqnarray}

Finally, using Equations \ref{T_u}, \ref{U_units} and \ref{Uxyz}, Eq.\ref{pc} can be written as follow,
\begin{equation}\label{opik_unit}
    \Gamma_{\rm opik} =\frac{1}{P} \frac{\sigma_{\rm opik}^2\,U_{\rm opik}}{\pi a_p^2\sin i_c\,U_x}\ ,
\end{equation}
where $P$ is the planet's orbital period.

%%%%%%%%%%%%%%%%%%%%%%%%%%%%%%%%%%%%%%%%%%%%%%%%
%% Tcoll vs Pcoll
%%%%%%%%%%%%%%%%%%%%%%%%%%%%%%%%%%%%%%%%%%%%%%%% 

\section{Comparison with The \"Opik Method}
\label{opikvscoll}

In this section, we compare our method for collision rates with the classic \"Opik theory to highlight the flexibility of our formulation. We choose two representative examples, comparing the rates behavior as a function of inclination ({\bf T-inc}) and eccentricity ({\bf T-ecc}) while keeping all other parameters constant. In both cases we choose the semi-major axis, eccentricity and true anomaly of the planet $a_p=5.2$\,au, $e_p=0.05$, and $f_c=0^\circ$ respectively, while for the particles we choose the values shown in Table\,\ref{table}. For consistency with \secref{nbody}, we used an inflated radius of collision, $R_{p}$=$(M_{p}/3M_\star)^{1/3}\,a_p$, where $M_\star$ is the mass of the host star, taken to be one solar mass. We note that the original formulation of $\Gamma_{\rm opik}$ (Eq. \ref{pc}) uses Jacobi normalized units. Therefore, in order to proper compare with $\Gamma_{\rm coll}$ (Eq. \ref{gamma}), we used $\Gamma_{\rm opik}$ with the proper units (Eq. \ref{opik_unit}).

\begin{table}[h!]
\centering
\caption{Input orbital elements of the particle: $a_c$, $e_c$, $i_c$, $\Omega_c$, $\omega_c$, and $f_c$, represents the semi-major axis, eccentricity, inclination, longitude of the ascending node, argument of periapsis, and true anomaly of the particle with respect to it's orbit about the star.}

\begin{tabular}{lc|ccccc} 
name & $e_c$ & $a_c$ & $i_c$ & $\Omega_c$& $\omega_c$ &$f_c$ \\
     & & [au] & [deg] & [deg] & [deg] &[deg]  \\
       \hline
{\bf T-inc}   & 0.1 & 6 & $0-180$ & 0 & 0 & 0  \\
 {\bf T-ecc}  & $0-1$ & 6 & $2.5$ & 0 & 0 & 0  \\
%\bottomrule 
\label{table}
 \end{tabular}
\end{table}

\begin{figure}
  \includegraphics[width=1.\columnwidth]{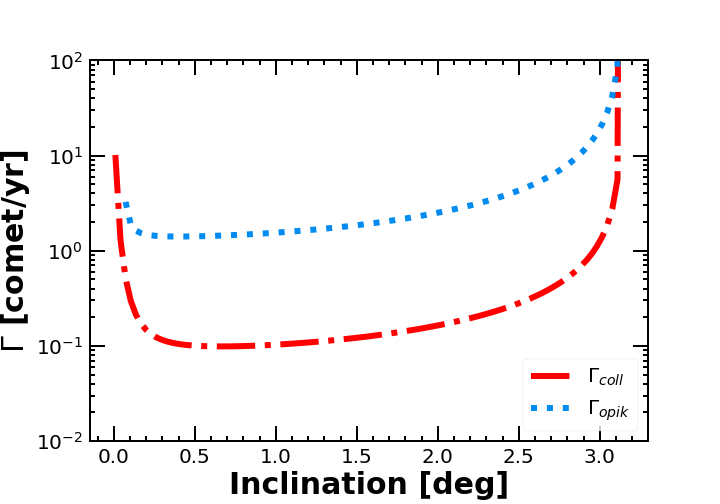}
  \includegraphics[width=1.\columnwidth]{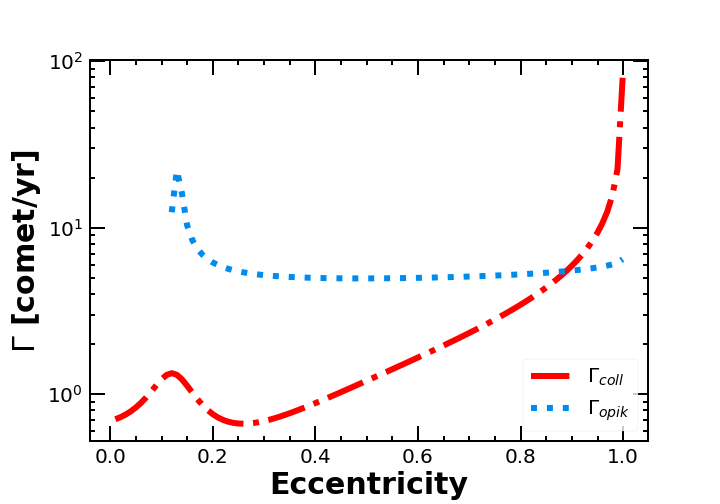}
  \caption{Compression  between  the $\Gamma_{\rm opik}$ (dotted blue line) and $\Gamma_{\rm coll}$ (dash-dot red line). The top panel shows the collision rate as a function of the inclination. While the bottom, the collision rate as a function of the eccentricity.}
  \label{tcollvspcoll}
\end{figure}

In \figref{tcollvspcoll} top panel, we show the collision rates as a function of the particle inclination with respect to the planet at the onset of collision for {\bf T-inc}. The two collision rates exhibit similar functional form as a function of the mutual inclination, $i$ since both are dominated by a similar volume dependency on the mutual inclination $V\sim \sin i$.

In \figref{tcollvspcoll} bottom panel, we show the collision rates as a function of the particle eccentricity at the onset of collision for  {\bf T-ecc}. As depicted in \figref{tcollvspcoll} bottom panel, our collision rate $\Gamma_{\rm coll}$ is qualitatively different from $\Gamma_{\rm opik}$. 
The \"Opik method predicts higher collision rates for eccentricities less than $\sim 0.85$, and it monotonically decreases from $e=0$ to $e=1$. In contrast, $\Gamma_{\rm coll}$ predicts an increasing distribution with two maximum. The first one for eccentricities within $0$ and $0.1$. While the second for $e_c$ within $0.4-1$.

%%%%%%%%%%%%%%%%%%%%%%%%%%%%%%%%%%%%%%%%%%%%%%%%%
%%%---------------------------- N-Body
\section{N-body simulations}
\label{sim}

\begin{figure*}
\centering
\includegraphics[width=1\textwidth]{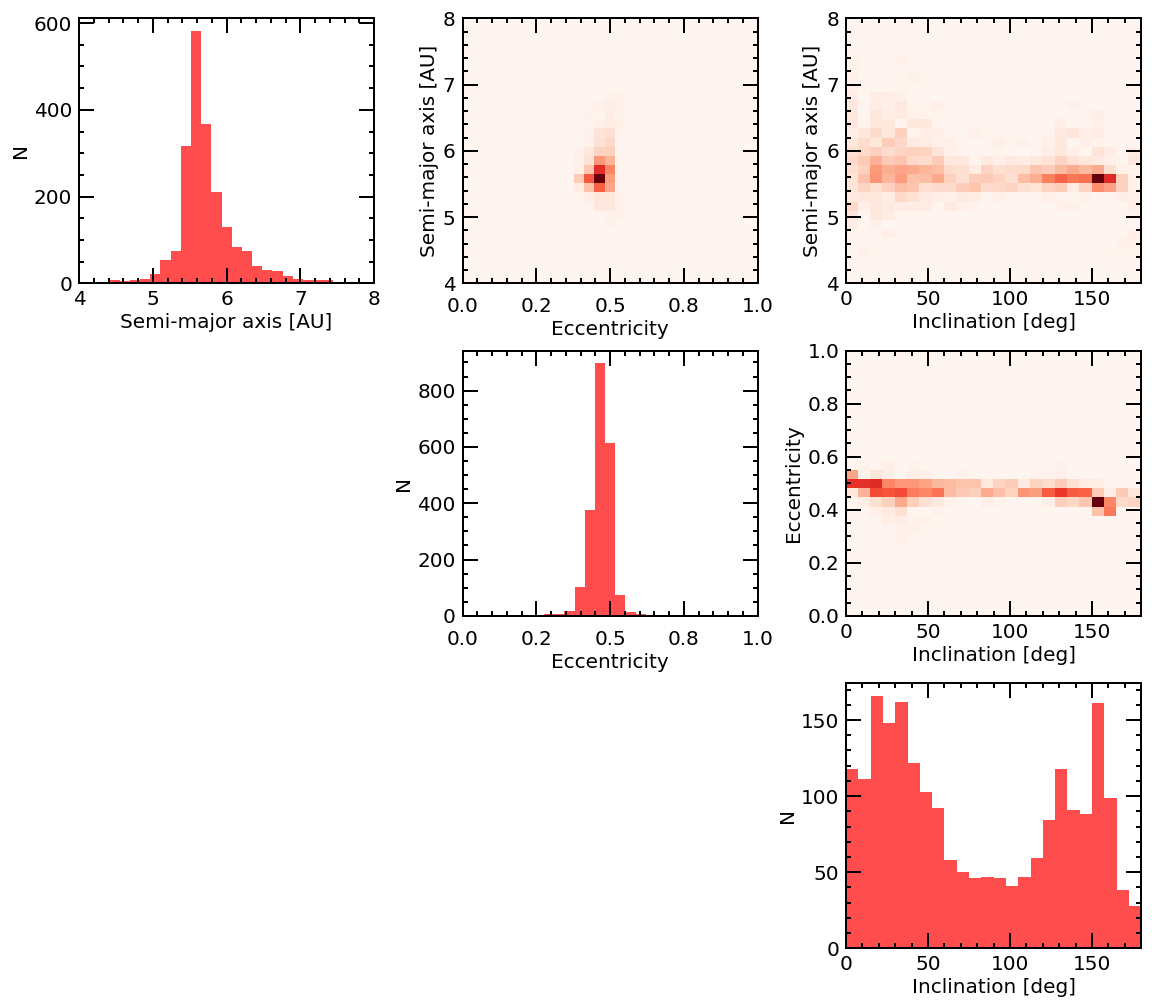}
 \caption{{\bf Run-inc}. Density map of semi-major axis, eccentricity and inclination of the collided particles.}
 \label{nbody_aei_inc}
\end{figure*}

\begin{figure*}
\centering
\includegraphics[width=1\textwidth]{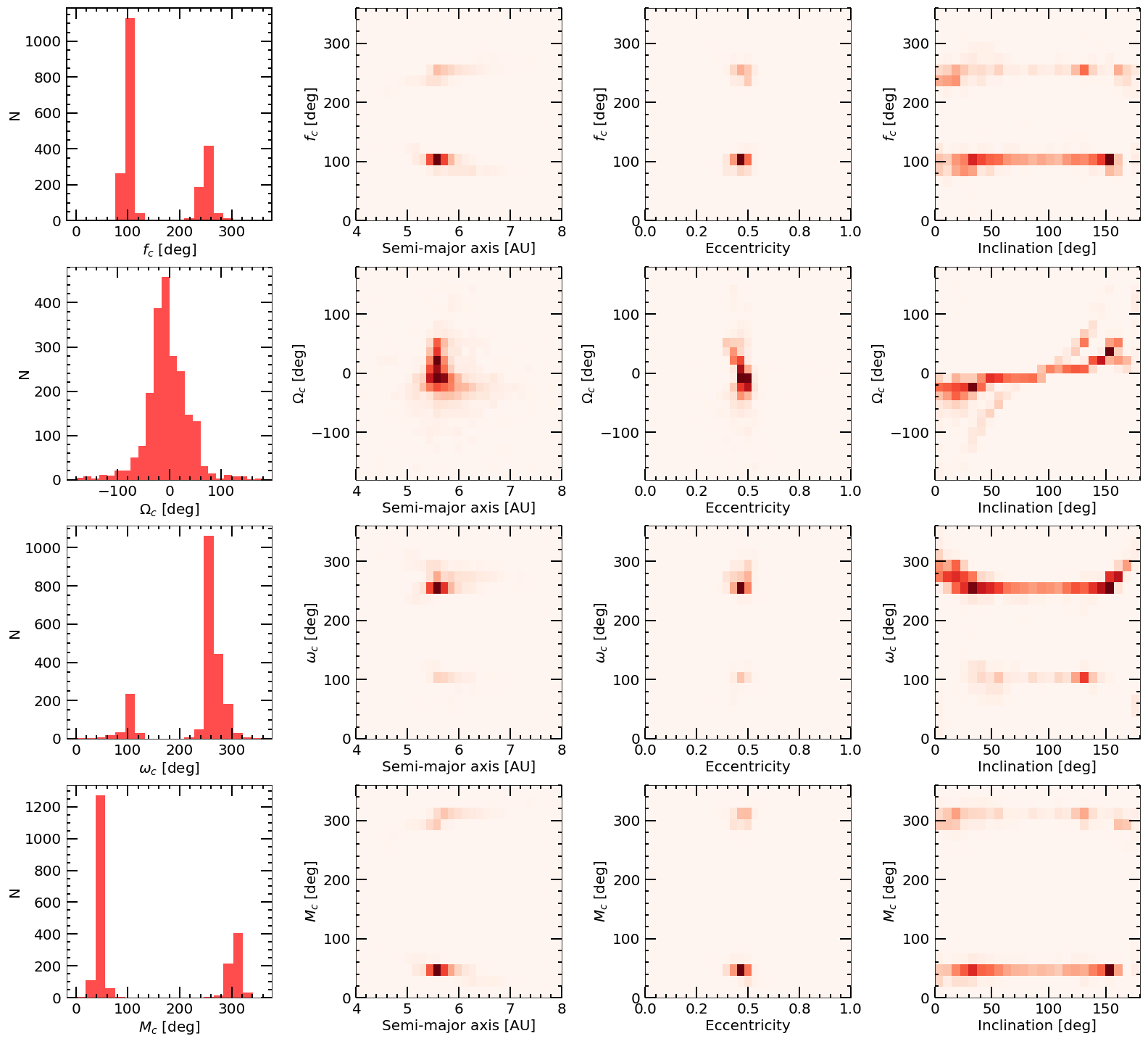}
 \caption{{\bf Run-inc}. Orbital elements density map of the collided particles. First row, shows the true anomaly $f_{c}$ as function of the semi-major axis $a_c$, eccentricity $e_c$ and inclination $i_c$. Second row, shows the longitude of the ascending node $\Omega_c$ as function of $a_c$, $e_c$, and $i_c$. Third row, shows the argument of periapsis $\omega_c$ as function of $a_c$, $e_c$, and $i_c$. Finally, last row shows the mean anomaly $M_c$ as function of $a_c$, $e_c$, and $i_c$.}
 \label{nbody_oe_inc}\end{figure*}

Following the method describe in Sect.\ref{nbody} and using
the input parameters shown in Table \ref{table1}, we preform two set of simulations, {\bf Run-inc} and {\bf Run-ecc}. In {\bf Run-inc} after $10,000$ years $\sim2,123$ particles collided with the planet ($\sim42.46\%$ of the initial particles). In \figref{nbody_aei_inc} we show the distribution of the collided particles, for the semi-major axis (first row), the eccentricity (second row) and the inclination (third row). We find that the particles with inclinations between $0-50$ and $130-160$ degrees are the most probable for collision. These particles have eccentricities within $0.4-0.6$ (see \figref{nbody_aei_inc} third column second panel). Additionally, the particles shown a bi-modal distribution in $f_c,\omega_c$ and $M_c$, having their maximums around $100^{\circ}$ and $250^{\circ}$ (\figref{nbody_oe_inc}). 
 
\begin{figure*}
\includegraphics[width=1.\textwidth]{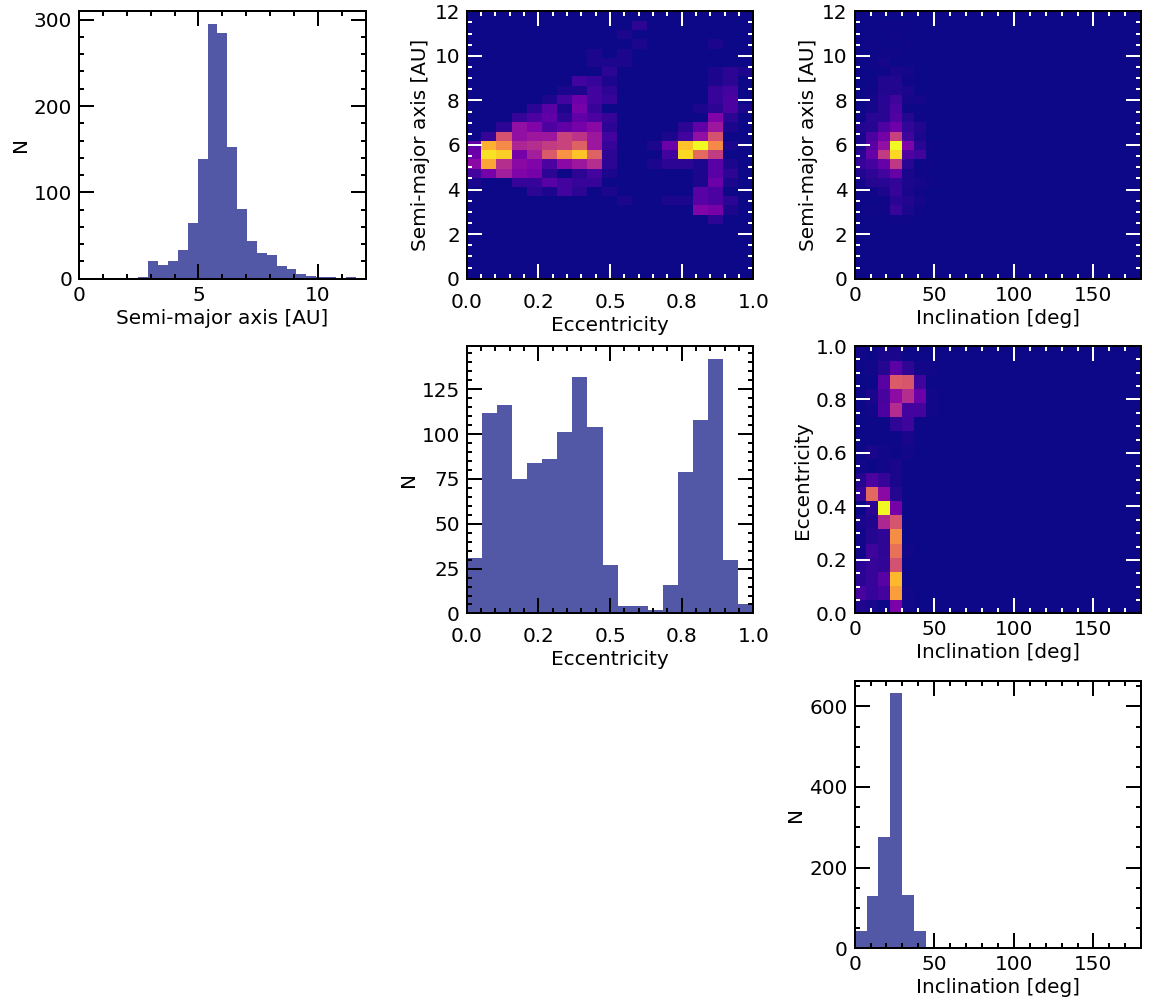}
 \caption{{\bf Run-ecc}. Density map of Semi-major axis, eccentricity and inclination of the collided particles.}
 \label{nbody_aei_ecc}
\end{figure*}

\begin{figure*}
\includegraphics[width=1.\textwidth]{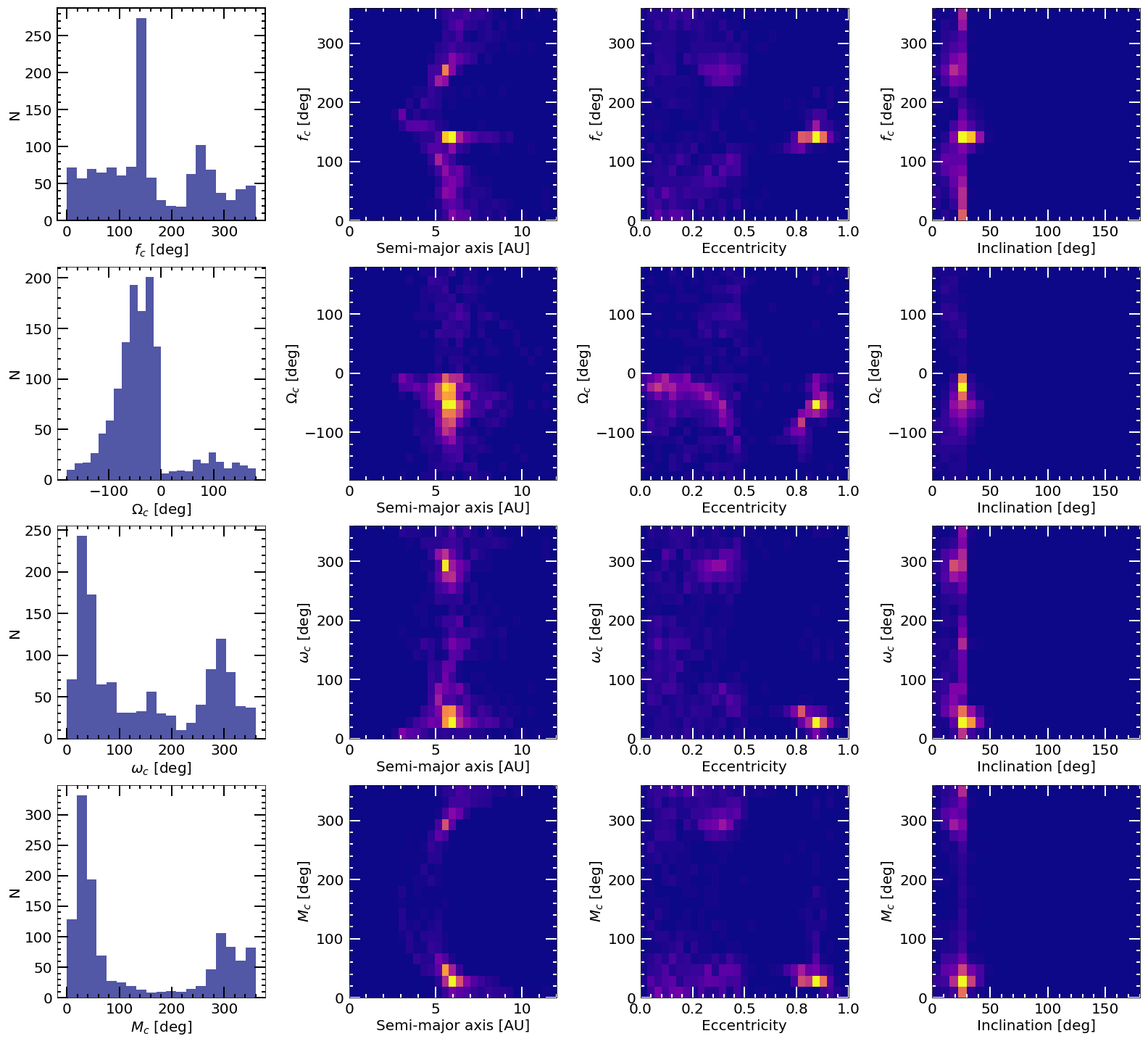}
 \caption{{\bf Run-ecc}. Orbital elements density map of the collided particles. First row, shows the true anomaly $f_{c}$ as function of the semi-major axis $a_c$, eccentricity $e_c$ and inclination $i_c$. Second row, shows the longitude of the ascending node $\Omega_c$ as function of $a_c$, $e_c$, and $i_c$. Third row, shows the argument of periapsis $\omega_c$ as function of $a_c$, $e_c$, and $i_c$. Finally, last row shows the mean anomaly $M_c$ as function of $a_c$, $e_c$, and $i_c$.}
 \label{nbody_oe_ecc}\end{figure*}

 In Figures \ref{nbody_aei_ecc} and \ref{nbody_oe_ecc} we show the results of the simulation {\bf Run-ecc}. We find that $\sim74.84\%$ of the particles remained in the system after $10,000$ years. The eccentricity of the collided particles ($\sim1,258$) formed a distribution with three peaks with maximums around $0.1$, $0.4$ and $0.7$ (second row \figref{nbody_aei_ecc}). Particles with eccentricity $\sim0.6$ did not collide. The longitude of the ascending node $\Omega_c$ of the particles has a preferred angle within $-100-0$ degrees (second-row \figref{nbody_oe_ecc}). These differed from the distribution of $\Omega_c$ in {\bf Run-inc}, where the particles have Gaussian distribution with maximum $\sim0$.  Overall, we find that the collided particles in {\bf Run-inc} and {\bf Run-ecc} have a strong dependency in the initial orbital elements, in particular the inclination and eccentricity.    

%%inc

%% ecc

%%%%%%%%%%%%%%%%%%%%%%%%%%%%%%%%%%%%%%%%%%%%%%%%%
% Don't change these lines
\bsp	% typesetting comment
\label{lastpage}
\end{document}